\documentclass[12pt,a4paper]{article}

\title{Two-charge rotating black holes in four-dimensional gauged supergravity}
\author{David D. K. Chow}
\date{}

\usepackage{amsmath}
\usepackage{amsfonts}
\usepackage{amssymb}
\usepackage{latexsym}

\makeatletter
\@addtoreset{equation}{section}
\makeatother

\newcommand{\be}{\begin{equation}}
\newcommand{\ee}{\end{equation}}
\newcommand{\ben}{\begin{equation}}
\newcommand{\een}{\end{equation}}
\newcommand{\bea}{\setlength\arraycolsep{2pt} \begin{eqnarray}}
\newcommand{\eea}{\end{eqnarray}}

\newcommand{\nnr}{\nonumber \\}
\newcommand{\lbl}{\label}

\newcommand{\eq}[1]{(\ref{#1})}
\newcommand{\se}{\section}
\newcommand{\sse}{\subsection}
\newcommand{\ssse}{\subsubsection}

\newcommand{\qd}{\quad}

\newcommand{\lt}{\left}
\newcommand{\rt}{\right}
\newcommand{\fr}{\frac}

\newcommand{\tf}{\tfrac}

\newcommand{\wtd}{\widetilde}

\newcommand{\df}{\textrm{d}}
\newcommand{\expe}[1]{\textrm{e}^{#1}}
\newcommand{\na}{\nabla}
\newcommand{\ol}{\overline}
\newcommand{\pd}{\partial}
\newcommand{\ra}{\rightarrow}

\newcommand{\sr}{\sqrt}

\newcommand{\ga}{\alpha}
\newcommand{\gb}{\beta}

\newcommand{\gd}{\delta}
\newcommand{\gD}{\Delta}
\newcommand{\gep}{\epsilon}

\newcommand{\gq}{\theta}

\newcommand{\gk}{\kappa}
\newcommand{\gl}{\lambda}
\newcommand{\gL}{\Lambda}

\newcommand{\gr}{\rho}

\newcommand{\gS}{\Sigma}
\newcommand{\gt}{\tau}

\newcommand{\gf}{\phi}
\newcommand{\gF}{\Phi}
\newcommand{\gvf}{\varphi}

\newcommand{\gw}{\omega}
\newcommand{\gW}{\Omega}

\newcommand{\uU}{\textrm{U}}

\newcommand{\im}{\textrm{i}}

\newcommand{\SL}{\textrm{SL}}
\newcommand{\SO}{\textrm{SO}}

\newcommand{\Tr}{\textrm{Tr}}

\newcommand{\va}{\textbf{a}}

\newcommand{\tp}{^\mathsf{T}}

\newcommand{\bbR}{\mathbb{R}}

\newcommand{\cF}{\mathcal{F}}

\newcommand{\cL}{\mathcal{L}}
\newcommand{\cM}{\mathcal{M}}
\newcommand{\cN}{\mathcal{N}}
\newcommand{\cO}{\mathcal{O}}

\newcommand{\ds}{\textrm{d} s^2}

\begin{document}

\thispagestyle{empty}

\begin{flushright}
MIFPA-10-54
\end{flushright}
\vspace*{100pt}
\begin{center}
{\bf \Large{Two-charge rotating black holes in four-dimensional gauged supergravity}}\\
\vspace{50pt}
\large{David D. K. Chow}
\end{center}

\begin{center}
% \textit{Department of Applied Mathematics and Theoretical Physics, University of Cambridge,\\
% Centre for Mathematical Sciences, Wilberforce Road, Cambridge CB3 0WA, UK}\\
\textit{George P. \& Cynthia W. Mitchell Institute for Fundamental Physics \& Astronomy,\\
Texas A\&M University, College Station, TX 77843-4242, USA}\\
%{\tt D.D.K.Chow@damtp.cam.ac.uk}\\
\texttt{chow@physics.tamu.edu}\\
\vspace{30pt}
{\bf Abstract\\}
\end{center}
We obtain an asymptotically AdS, non-extremal, electrically charged and rotating black hole solution of 4-dimensional $\uU (1)^4$ gauged supergravity with 2 non-zero and independent $\uU (1)$ charges.  The thermodynamical quantities are computed.  We find BPS solutions that are nakedly singular.  The solution is generalized to include a NUT parameter and dyonic gauge fields.  The string frame metric has a rank-2 Killing--St\"{a}ckel tensor and has completely integrable geodesic motion, and the massless Klein--Gordon equation separates for the Einstein frame metric.

\newpage

%%%%%%%%%%%%%%%%%

\se{Introduction}

%%%%%%%%%%%%%%%%%

The AdS/CFT correspondence has motivated the construction of asymptotically AdS black holes, which may be understood in terms of a boundary field theory.  In particular, we know many examples of AdS black hole solutions of gauged supergravity theories in a variety of dimensions; see \cite{emprea, chow} for a review.  Supersymmetric black holes are of special interest, because then the correspondence is on a firmer footing.  However, supersymmetric AdS black holes must rotate, and including rotation makes the construction of solutions considerably more complicated.  More generally, non-extremal AdS black holes are also of interest for understanding the AdS/CFT correspondence at non-zero temperature.

The construction of charged and rotating AdS black holes in gauged supergravities with the maximum number of independent angular momenta and $\uU (1)$ charges remains an open problem.  Restricting to abelian gauge fields, the relevant theories in spacetime dimensions $D = 4, 5, 7$ are have respective gauge groups $\uU (1)^4$, $\uU (1)^3$, and $\uU (1)^2$.  In gauged supergravity, unlike ungauged supergravity, there are no solution generating techniques available for the construction of charged and rotating black holes.  Therefore, constructing these black holes requires guesswork, inspired by the structure of known solutions.

To better understand their structures, a most basic example of an asymptotically AdS, charged and rotating black hole in gauged supergravity was recently constructed \cite{chow0}.  This is a solution of 4-dimensional $\uU (1)^4$ gauged supergravity with only 1 non-zero $\uU (1)$ charge.  The other known asymptotically AdS, charged and rotating black hole solutions of this theory are the Kerr--Newman--AdS solution \cite{carter0, carter}, which has all 4 charges equal, and a more general solution with the 4 charges pairwise equal \cite{chcvlupo}.  In this paper, we continue to take advantage of the simplicity of this low dimension and obtain an AdS black hole solution of 4-dimensional $\uU (1)^4$ gauged supergravity with 2 non-zero and independent $\uU (1)$ charges.

However, the motivation for constructing this 2-charge solution extends beyond 4 dimensions.  For arbitrary dimension, there is the 2-charge Cveti\v{c}--Youm solution \cite{cveyou3}, which represents a rotating black hole with the maximum number of independent angular momenta and 2 non-zero $\uU (1)$ charges.  In particular, it is a solution of ungauged supergravity in $D = 4, 5, 7$.  If both $\uU (1)$ charges are equal, then the solution simplifies considerably \cite{chow3}.  The equal charge simplification persists with the inclusion of gauging, regardless of dimension.  These equal-charge AdS black holes in $D = 4$ \cite{chcvlupo}, $D = 5$ \cite{chcvlupo2}, $D = 7$ \cite{chow3}, and also a solution in $D = 6$ \cite{chow2}, can be written in a fairly unified manner.  We therefore expect that the AdS black hole with 2 independent $\uU (1)$ charges in $D = 4$ will be a useful guide for constructing AdS black holes with 2 independent $\uU (1)$ charges in $D = 5$ and $D = 7$.

The outline of this paper is as follows.  In Section 2, we present a charged and rotating solution of 4-dimensional $\uU (1)^4$ gauged supergravity with 2 independent and non-zero $\uU (1)$ charges and 2 zero $\uU (1)$ charges.  We compute the thermodynamic quantities.  Although there are BPS solutions, they are nakedly singular.  In Section 3, we generalize by including a NUT parameter and allowing for magnetic charge, obtaining dyonic solutions.  In Section 4, we present a rank-2 Killing--St\"{a}ckel tensor for the string frame metric.  We demonstrate explicitly the separability of the Hamilton--Jacobi equation for geodesic motion in the string frame metric and the massless Klein--Gordon equation in the Einstein frame metric.  We conclude in Section 5.

%%%%%%%%%%%%%%%%%%%%%%%%%%%%%%%%%%%%%%

\se{Kerr--AdS with 2 electric charges}

%%%%%%%%%%%%%%%%%%%%%%%%%%%%%%%%%%%%%%

The Kerr--AdS metric \cite{carter0, carter} is an uncharged, rotating black hole solution of Einstein gravity with a cosmological constant.  We shall generalize it as a solution of 4-dimensional $\uU (1)^4$ gauged supergravity by including 2 non-zero and independent $\uU (1)$ charges, first allowing only electric, not magnetic, charge.  Then, we shall study the thermodynamical quantities and BPS solutions.

%%%%%%%%%%%%%%%%%%%%%%%%%%%%%%%%%%%%%

\sse{$\uU (1)^4$ gauged supergravity}

%%%%%%%%%%%%%%%%%%%%%%%%%%%%%%%%%%%%%

The maximal 4-dimensional $\cN = 8$, $\SO (8)$ gauged supergravity can be consistently truncated to $\uU (1)^4$ gauged supergravity, which is $\cN = 2$ gauged supergravity coupled to 3 abelian vector multiplets.  The bosonic fields of this truncation are the graviton $g_{a b}$, 4 $\uU (1)$ gauge fields $A_{(1)}^I$, 3 dilatons $\gvf_i$ and 3 axions $\chi_i$, with $I = 1, 2, 3, 4$ and $i = 1, 2, 3$.  The bosonic Lagrangian, which was given in \cite{cveticetal}, can be written as
%%%
\bea
\cL & = & R \star 1 - \fr{1}{2} \sum_{i = 1}^3 ( \star \df \gvf_i \wedge \df \gvf_i + \expe{- 2 \gvf_i} \star \df \chi_i \wedge \df \chi_i ) - \fr{1}{2} \sum_{I = 1}^4 X_I^{-2} \star F_{(2)}^I \wedge F_{(2)}^I \nnr
&& + \chi_1 (F_{(2)}^1 \wedge F_{(2)}^2 + F_{(2)}^3 \wedge F_{(2)}^4 ) - V \star 1 . \lbl{Lagrangian1}
\eea
%%%
The 2-form field strengths are
%%%
\bea
F_{(2)}^1 & = & \df A_{(1)}^1 , \nnr
F_{(2)}^2 & = & \df A_{(1)}^2 + \chi_2 \, \df A_{(1)}^3 - \chi_3 \, \df A_{(1)}^4 + \chi_2 \chi_3 \, \df A_{(1)}^1 , \nnr
F_{(2)}^3 & = & \df A_{(1)}^3 + \chi_3 \, \df A_{(1)}^1 , \nnr
F_{(2)}^4 & = & \df A_{(1)}^4 - \chi_2 \, \df A_{(1)}^1 ,
\eea
%%%
and we have used the dilaton combinations
%%%
\ben
X_I = \exp ( - \tf{1}{2} \va_I \cdot \boldsymbol{\gvf} ) , \qd \boldsymbol{\gvf} = (\gvf_1 , \gvf_2 , \gvf_3) ,
\een
%%%
with
%%%
\ben
\va_1 = (1, 1, 1) , \qd \va_2 = (1, -1, -1) , \qd \va_3 = (-1, 1, -1) , \qd \va_4 = (-1, -1, 1) .
\een
%%%
The potential is
%%%
\ben
V = - g^2 \sum_{i = 1}^3 (2 \cosh \gvf_i + \chi_i^2 \expe{- \gvf_i}) .
\een
%%%

We shall consider black hole solutions with 2 non-zero $\uU (1)$ charges, and so truncate further by taking $A_{(1)}^3 = A_{(1)}^4 = 0$, $\gvf_2 = \gvf_3$ and $\chi_2 = \chi_3 = 0$, denoting $\chi = \chi_1$.  The remaining bosonic Lagrangian becomes
%%%
\bea
\cL & = & R \star 1 - \fr{1}{2} \star \df \gvf_1 \wedge \df \gvf_1 - \fr{1}{2} \expe{- 2 \gvf_1} \star \df \chi \wedge \df \chi - \star \df \gvf_2 \wedge \df \gvf_2 - \fr{1}{2} \sum_{I = 1}^2 X_I^{-2} \star F_{(2)}^I \wedge F_{(2)}^I \nnr
&& + \chi F_{(2)}^1 \wedge F_{(2)}^2 - V \star 1 , \lbl{Lagrangian2}
\eea
%%%
where $F_{(2)}^I = \df A_{(1)}^I$, with $I = 1, 2$, and now
%%%
\bea
&& X_1 = \expe{- \gvf_1 / 2 - \gvf_2} , \qd X_2 = \expe{- \gvf_1 / 2 + \gvf_2} , \nnr
&& V = - g^2 (\expe{\gvf_1} + \expe{- \gvf_1} + 2 \expe{\gvf_2} + 2 \expe{- \gvf_2} + \chi^2 \expe{- \gvf_1}) . 
\eea
%%%
The resulting field equations are
%%%
\bea
G_{a b} & = & \tf{1}{2} \na_a \gvf_1 \, \na_b \gvf_1 - \tf{1}{4} \na^c \gvf_1 \, \na_c \gvf_1 \, g_{a b} + 2 (\tf{1}{2} \na_a \gvf_2 \, \na_b \gvf_2 - \tf{1}{4} \na^c \gvf_2 \, \na_c \gvf_2 \, g_{a b}) \nnr
&& + \expe{\gvf_1 + 2 \gvf_2} (\tf{1}{2} F{^1}{_a}{^c} F{^1}{_{b c}} - \tf{1}{8} F{^1}{^{c d}} F{^1}{_{c d}} g_{a b}) + \expe{\gvf_1 - 2 \gvf_2} (\tf{1}{2} F{^2}{_a}{^c} F{^2}{_{b c}} - \tf{1}{8} F{^2}{^{c d}} F{^2}{_{c d}} g_{a b}) \nnr
&& + \expe{- 2 \gvf_1} (\tf{1}{2} \na_a \chi \, \na_b \chi - \tf{1}{4} \na^c \chi \, \na_c \chi \, g_{a b}) + \tf{1}{2} g^2 (\expe{\gvf_1} + \expe{- \gvf_1} + 2 \expe{\gvf_2} + 2 \expe{- \gvf_2} + \chi^2 \expe{- \gvf_1}) g_{a b} , \nnr
\eea
%%%
and
%%%
%%%
\bea
&& \na_a (\expe{\gvf_1 + 2 \gvf_2} F^{1 a b}) = \tf{1}{2} \gep^{b c d e} (\na_c \chi) F{^2}{_{d e}} , \nnr
&& \na_a (\expe{\gvf_1 - 2 \gvf_2} F^{2 a b}) = \tf{1}{2} \gep^{b c d e} (\na_c \chi) F{^1}{_{d e}} , \nnr
&& \square \gvf_1 = \tf{1}{4} \expe{\gvf_1 + 2 \gvf_2} F^{1 a b} F{^1}{_{a b}} + \tf{1}{4} \expe{\gvf_1 - 2 \gvf_2} F^{2 a b} F{^2}{_{a b}} - \expe{- 2 \gvf_1} \na^a \chi \, \na_a \chi - g^2 (\expe{\gvf_1} - \expe{- \gvf_1} - \chi^2 \expe{- \gvf_1}) , \nnr
&& \square \gvf_2 = \tf{1}{4} \expe{\gvf_1 + 2 \gvf_2} F^{1 a b} F{^1}{_{a b}} - \tf{1}{4} \expe{\gvf_1 - 2 \gvf_2} F^{2 a b} F{^2}{_{a b}} - g^2 (\expe{\gvf_2} - \expe{- \gvf_2}) , \nnr
&& \na_a (\expe{- 2 \gvf_1} \na^a \chi) = \tf{1}{4} \gep^{a b c d} F{^1}{_{a b}} F{^2}{_{c d}} - 2 g^2 \chi \expe{- \gvf_1} .
\eea
%%%
From the field equations for $F_{(2)}^I$, we introduce the dual field strengths and potentials
%%%
\ben
\df \wtd{A}_{(1)}^1 = \wtd{F}_{(2)}^1 = X_1^{-2} \star F_{(2)}^1 - \chi F_{(2)}^2 , \qd \df \wtd{A}_{(1)}^2 = \wtd{F}_{(2)}^2 = X_2^{-2} \star F_{(2)}^2 - \chi F_{(2)}^1 . \lbl{dual}
\een
%%%

%%%%%%%%%%%%%%%%%%%%%%%%%

\sse{Black hole solution}

%%%%%%%%%%%%%%%%%%%%%%%%%

A rotating black hole solution with 2 independent $\uU (1)$ charges is
%%%
\bea
\ds & = & \fr{1}{\sr{H_1 H_2} \gr^2 \Xi^2} \bigg( - (\gD_\gq \gD_r - V_r^2 a^2 \sin^2 \gq) \gD_\gq \, \df t^2 + (\gD_\gq \wtd{V}_r^2 a^2 - \gD_r \sin^2 \gq) \, a^2 \sin^2 \gq \, \df \gf^2 \nnr
&& - 2 m a r c_1 c_2 \wtd{c}_1 \wtd{c}_2 \gD_\gq \sin^2 \gq \, 2 \, \df t \, \df \gf \bigg) + \sr{H_1 H_2} \lt( \fr{\gr^2}{\gD_r} \, \df r^2 + \fr{\gr^2}{\gD_\gq} \, \df \gq^2 \rt) , \nnr
A_{(1)}^1 & = & \fr{2 m r s_1}{H_1 \gr^2} \bigg( c_1 \wtd{c}_2 \gD_\gq \, \fr{\df t}{\Xi} - a c_2 \wtd{c}_1 \sin^2 \gq \, \fr{\df \gf}{\Xi} \bigg) , \qd A_{(1)}^2 = \fr{2 m r s_2}{H_2 \gr^2} \bigg( c_2 \wtd{c}_1 \gD_\gq \, \fr{\df t}{\Xi} - a c_1 \wtd{c}_2 \sin^2 \gq \, \fr{\df \gf}{\Xi} \bigg) , \nnr
X_1 & = & H_1^{-3/4} H_2^{1/4} , \qd X_2 = H_1^{1/4} H_2^{-3/4} , \qd \chi = \fr{2 m a s_1 s_2 \cos \gq}{\gr^2} ,
\eea
%%%
where
%%%
\bea
&& \gD_r = r^2 + a^2 - 2 m r + g^2 r^2 r_1 r_2 + a^2 g^2 r^2 - 2 m r a^2 g^2 s_1^2 s_2^2 , \qd \gD_\gq = 1 - a^2 g^2 \cos^2 \gq , \nnr
&& V_r^2 = (1 + g^2 r r_1) (1 + g^2 r r_2) , \qd \wtd{V}_r^2 = (1 + r r_1 / a^2) (1 + r r_2 / a^2) , \qd r_I = r + 2 m s_I^2 , \nnr
&& H_I = 1 + \fr{2 m r s_I^2}{\gr^2} , \qd s_I = \sinh \gd_I , \qd c_I = \cosh \gd_I , \qd \wtd{c}_I = \sr{1 + a^2 g^2 s_I^2} , \nnr
&& \gr^2 = r^2 + a^2 \cos^2 \gq , \qd \Xi = 1 - a^2 g^2 .
\eea
%%%

We have presented the solution using asymptotically static Boyer--Lindquist-type coordinates.  The azimuthal coordinate $\gf$ has canonical normalization, with period $2 \pi$.  The temporal coordinate $t$ also has canonical normalization, fixed because the spacetime is asymptotically AdS.  If $m = 0$, then the coordinate change
%%%
\ben
\Xi \hat{r}^2 \sin^2 \hat{\gq} = (r^2 + a^2) \sin^2 \gq , \qd \hat{r}^2 \cos^2 \hat{\gq} = r^2 \cos^2 \gq ,
\een
%%%
gives anti-de Sitter spacetime in the canonical form
%%%
\ben
\ds = - (1 + g^2 \hat{r}^2) \, \df t^2 + \fr{\df \hat{r}^2}{1 + g^2 \hat{r}^2} + \hat{r}^2 \, \df \hat{\gq}^2 + \sin^2 \hat{\gq} \, \df \gf^2 .
\een
%%%

The solution has 5 parameters: a mass parameter $m$; a rotation parameter $a$; 2 charge parameters, $\gd_1$ and $\gd_2$; and a gauge-coupling constant $g$.  Without rotation, when $a = 0$, the solution reduces to a particular case of the 4-charge static solution \cite{dufliu}, but with only 2 of the 4 charges non-zero.  Without any charge, when $\gd_1 = \gd_2 = 0$, the solution reduces to the 4-dimensional Kerr--AdS metric \cite{carter0, carter}.  With only 1 non-zero charge, say when $\gd_2 = 0$, the solution reduces to the recently discovered 1-charge solution \cite{chow0}.  With both charges equal, when $\gd_1 = \gd_2$, the solution reduces to the gauged solution of \cite{chcvlupo}, where the 4 charges are pairwise equal, but here with one pair of charges zero and one pair non-zero; we shall give the link more precisely later.

Without gauging, when $g = 0$, the solution reduces to a 2-charge solution of Sen \cite{sen}, up to global transformations of scalars and vectors.  Note that Sen uses parameters $\ga$ and $\gb$ that are related to our $\gd_1$ and $\gd_2$ by $\ga = \gd_1 + \gd_2$ and $\gb = \gd_1 - \gd_2$.  The ungauged solution can also be regarded as the 4-dimensional case of the 2-charge Cveti\v{c}--Youm solution \cite{cveyou3}, or the 4-charge Cveti\v{c}--Youm solution \cite{cveyou, chcvlupo}, but with only 2 of the 4 charges non-zero.  A 3-charge ungauged solution \cite{jamupa} uses the Sen parameterization for 2 charge parameters, and our Cveti\v{c}--Youm parameterization for the third charge.  With only 1 non-zero charge, say when $\gd_2 = 0$, the ungauged solution reduces to the Frolov--Zelnikov--Bleyer solution \cite{frzebl}, presented there with $v = \tanh \gd_1$.

To discover this new solution, we are helped by the structure of these limits, most significantly by the 1-charge solution \cite{chow0}.  One helpful feature is that the solution is invariant under the transformation
%%%
\bea
a \ra \fr{1}{a g^2} , \qd r \ra \fr{r}{a g} , \qd y \ra \fr{y}{a g} , \qd m \ra \fr{m}{a^3 g^3} , \qd \gf \ra g t , \qd t \ra \fr{\gf}{g} , \qd s_I \ra a g s_I , \lbl{inversion}
\eea
%%%
where $y = a \cos \gq$.  This is a discrete inversion symmetry, under which the rotation parameter $a$ is inverted through the AdS radius $1 / g$.  It is noteworthy that the symmetry interchanges $g t$ and $\gf$ and interchanges $c_I$ and $\wtd{c}_I$.  The symmetry interchanges over-rotating solutions, with $| a g | > 1$, and under-rotating solutions, with $| a g | < 1$.  This inversion symmetry was first found in \cite{chlupo3} for the 5-dimensional Kerr--AdS solution \cite{hahuta}, then for the Kerr--NUT--AdS solutions of Einstein gravity in arbitrary dimension \cite{chlupo2}, and more recently for the 4-dimensional 1-charge solution \cite{chow0}.

4-dimensional $\cN = 8$, $\SO (8)$ gauged supergravity is a consistent dimensional reduction of 11-dimensional supergravity on $S^7$ \cite{dewnic}.  In principle, a solution of the 4-dimensional theory can be uplifted to a solution of the 11-dimensional theory, but in practice the general consistency proof is rather implicit.  The explicit reduction ansatz is not known for the full 4-dimensional $\uU (1)^4$ gauged supergravity, and so it is not known how to embed our general 2-charge solution into 11 dimensions.  However, the reduction ansatz is known for two cases of relevance here.  One relevant case is if the axions vanish \cite{cveticetal}; this suffices for embedding into 11 dimensions the 4-charge static solution \cite{dufliu} and the 1-charge rotating solution \cite{chow0}.  The second relevant case is $\cN = 4$, $\SO (4)$ gauged supergravity \cite{cvlupo}; this suffices for embedding into 11 dimensions the rotating solution where the 4 charges are pairwise equal \cite{chcvlupo}, which includes taking $\gd_1 = \gd_2$ in our 2-charge solution.

%%%%%%%%%%%%%%%%%%%%

\sse{Thermodynamics}

%%%%%%%%%%%%%%%%%%%%

The outer black hole horizon is located at the largest root of $\gD_r (r)$, say at $r = r_+$.  By assumption, the parameters of the solution are chosen so that we have a black hole rather than a naked singularity, so such a positive root exists.  The angular velocity $\gW$ is constant over the horizon and is obtained from the Killing vector
%%%
\ben
l = \fr{\pd}{\pd t} + \gW \fr{\pd}{\pd \gf}
\een
%%%
that becomes null on the horizon.  The electrostatic potentials $\gF_I$ and surface gravity $\gk$ are also constant over the horizon.  They are given by evaluating on the horizon both $\gF_I = l \cdot A_{(1)}^I$ and $l^b \na_b l^a = \gk l^a$, and then the Hawking temperature is $T = \gk / 2 \pi$.  The angular momentum and electric charges are given by the respective Komar integrals
%%%
\ben
J = \fr{1}{16 \pi} \int _{S_\infty^2} \! \star \df K , \qd Q_I = \fr{1}{16 \pi} \int_{S_\infty^2} \! \wtd{F}_{(2)}^I ,
\een
%%%
where $K = K_a \, \df x^a$ is the 1-form obtained from the rotational Killing vector $K^a \, \pd_a = \pd / \pd \gf$.  The horizon area $A$ is obtained by integrating the square root of the determinant of the induced metric on a time slice of the horizon, and then the Bekenstein--Hawking entropy is $S = A / 4$.

One finds that $T \, \df S + \gW \, \df J + \gF_1 \, \df Q_1 + \gF_2 \, \df Q_2$ is an exact differential, and so, as advocated by \cite{gipepo}, we may integrate the first law of black hole mechanics,
%%%
\ben
\df E = T \, \df S + \gW \, \df J + \gF_1 \, \df Q_1 + \gF_2 \, \df Q_2 ,
\een
%%%
to obtain the thermodynamic mass $E$.  There are also several other definitions of mass for asymptotically AdS spacetimes in the literature; one is the AMD (Ashtekar--Magnon--Das) mass, defined for 4 dimensions \cite{ashmag} and higher \cite{ashdas}.  One introduces a conformally rescaled metric $\ol{g}_{a b} = \gW^2 g_{a b}$ such that on the conformal boundary both $\gW = 0$ and $\df \gW \neq 0$.  Its Weyl tensor is $\ol{C}{^a}{_{b c d}}$, and we define $\ol{n}_a = \pd_a \gW$.  For an asymptotic Killing vector field $K$, which here is $K = \pd / \pd t$, there is an associated conserved quantity.  In 4 dimensions the AMD mass is
%%%
\ben
E = \fr{1}{8 \pi g^3} \int_\gS \! \df \ol{\gS}_a \, \gW^{-1} \ol{n}^c \ol{n}^d \ol{C}{^a}{_{c b d}} K^b , \lbl{AMDmass}
\een
%%%
where $\df \ol{\gS}_a$ is the area element of the $S^2$ section of the conformal boundary.  See \cite{gipepo, chlupo} for many AMD mass computations for AdS black holes.  For our solution, we take $\gW = 1 / g r$ for definiteness.  As $r \ra \infty$, the Weyl tensor component $C{^t}{_{r t r}}$ behaves as\footnote{We correct a typographical error in \cite{chow0} here.}
%%%
\ben
C{^t}{_{r t r}} = \fr{m}{\Xi g^2 r^5} (2 \Xi + 3 a^2 g^2 \sin^2 \gq) [1 + \tf{1}{2} (1 + a^2 g^2) (s_1^2 + s_2^2) + a^2 g^2 s_1^2 s_2^2] + \cO \lt( \fr{1}{r^6} \rt) .
\een
%%%
The conformal boundary has metric
%%%
\ben
\df \ol{s}_3^2 = - \fr{\gD_\gq}{\Xi} \, \df t^2 + \fr{1}{g^2 \gD_\gq} \, \df \gq^2 + \fr{\sin^2 \gq}{\Xi g^2} \, \df \gf^2 .
\een
%%%
Substituting these into \eq{AMDmass}, we find that the AMD mass agrees with the thermodynamic mass.

In summary, we find the thermodynamic quantities
%%%
\bea
E & = & \fr{m}{\Xi^2} [1 + \tf{1}{2} (1 + a^2 g^2) (s_1^2 + s_2^2) + a^2 g^2 s_1^2 s_2^2] , \nnr
S & = & \fr{\pi \sr{(r_+^2 + a^2 + 2 m s_1^2 r_+) (r_+^2 + a^2 + 2 m s_2^2 r_+)}}{\Xi} , \nnr
T & = & \fr{r_+^2 - a^2 + g^2 r_+^2 (3 r_+^2 + 4 m s_1^2 r_+ + 4 m s_2^2 r_+ + 4 m^2 s_1^2 s_2^2 + a^2)}{4 \pi r_+ \sr{(r_+^2 + a^2 + 2 m s_1^2 r_+) (r_+^2 + a^2 + 2 m s_2^2 r_+)}} , \nnr
J & = & \fr{c_1 c_2 \wtd{c}_1 \wtd{c}_2 m a}{\Xi^2} , \qd \gW = \fr{a [1 + g^2 r_+ (r_+ + 2 m s_1^2)] [1 + g^2 r_+ (r_+ + 2 m s_2^2)]}{2 m r_+ c_1 c_2 \wtd{c}_1 \wtd{c}_2} , \nnr
Q_1 & = & \fr{m s_1 c_1 \wtd{c}_2}{2 \Xi} , \qd \gF_1 = \fr{2 m s_1 c_1 \wtd{c}_2 r_+}{r_+^2 + a^2 + 2 m s_1^2 r_+} , \nnr
Q_2 & = & \fr{m s_2 c_2 \wtd{c}_1}{2 \Xi} , \qd \gF_2 = \fr{2 m s_2 c_2 \wtd{c}_1 r_+}{r_+^2 + a^2 + 2 m s_2^2 r_+} .
\lbl{thermo}
\eea
%%%

%%%%%%%%%%%%%%%%%%%%%%%%%%%%%%

\sse{Supersymmetric solutions}

%%%%%%%%%%%%%%%%%%%%%%%%%%%%%%

4-dimensional $\uU (1)^4$ gauged supergravity is known to have supersymmetric AdS$_4$ black holes.  With 4 equal $\uU (1)$ charges, they are known \cite{kosper, calkle} amongst the Kerr--Newman--AdS family of solutions \cite{carter0, carter}.  With pairwise equal charges, they are known \cite{cvgilupo} amongst the solutions of \cite{chcvlupo}.

By considering eigenvalues of the Bogomolny matrix, the BPS condition, up to a choice of signs, is \cite{cvgilupo}
%%%
\ben
E - g J - Q_1 - Q_2 = 0 . \lbl{BPScondition}
\een
%%%
Expressed in terms of the parameters of our solution, this leads to
%%%
\ben
c_1^2 \wtd{c}_2^2 + c_2^2 \wtd{c}_1^2 - (1 - a^2 g^2) (s_1 c_1 \wtd{c}_2 + s_2 c_2 \wtd{c}_1) - 2 a g c_1 c_2 \wtd{c}_1 \wtd{c}_2 = 0 .
\een
%%%
This is solved by
%%%
\ben
a g s_1 s_2 = 1 , \lbl{BPS}
\een
%%%
provided that $\gd_1$ and $\gd_2$ are positive; different signs for $\gd_I$ would also satisfy BPS conditions, but with some sign changes in \eq{BPScondition}.  Note that there are no BPS solutions if 1 of the 2 charges vanishes.

For a BPS solution, the radial function of the metric becomes
%%%
\ben
R = \lt( 2 m s_1 s_2 g r - \fr{1}{g s_1 s_2} \rt) ^2 + g^2 r^4 + 2 m (s_1^2 + s_2^2) g^2 r^3 + \lt( 1 + \fr{1}{s_1^2 s_2^2} \rt) r^2 .
\een
%%%
Because $R$ is a sum of non-negative terms, it generically has no real roots, and so the solution has a naked singularity.  If there is a root, then it is at $r = r_+$, with $r_+ = 1 / 2 m s_1^2 s_2^2 g^2$.  However, substitution back into $R$ gives
%%%
\ben
R | _{r = r_+} = \fr{1}{4 m^2 g^4 s_1^6 s_2^6} \lt( \fr{1}{4 m^2 g^2 s_1^2 s_2^2} + c_1^2 c_2^2 \rt) ,
\een
%%%
which cannot vanish.  Therefore, our 2-charge solution does not include supersymmetric AdS$_4$ black holes.  Note the similarity with 5-dimensional $\uU (1)^3$ gauged supergravity, where there are no supersymmetric AdS$_5$ black holes with 2 of the 3 $\uU (1)$ charges zero, although there are with only 1 of the 3 charges zero \cite{kulure}.  Also, in 7-dimensional $\uU (1)^2$ gauged supergravity, if the rotation parameters are all equal, then there are no supersymmetric AdS$_7$ black holes with 1 of the 2 $\uU (1)$ charges zero \cite{cvgilupo}.

%%%%%%%%%%%%%%%%%%%%

\se{Generalizations}

%%%%%%%%%%%%%%%%%%%%

We shall generalize our solution in two ways.  Firstly, we shall include a NUT parameter.  Secondly, we shall allow for magnetic charge, using global symmetries to generate dyonic solutions.

%%%%%%%%%%%%%%%%%%%

\sse{NUT parameter}

%%%%%%%%%%%%%%%%%%%

A more general solution that includes a NUT parameter $\ell$ is
%%%
\bea
\ds & = & \fr{1}{\sr{H_1 H_2} (r^2 + y^2) \Xi^2} \bigg( - (V_y^2 R - V_r^2 Y) \, \df t^2 + (\wtd{V}_r^2 Y - \wtd{V}_y^2 R) \, a^2 \, \df \gf^2 \nnr
&& - \fr{2 c_1 c_2 \wtd{c}_1 \wtd{c}_2 (m r Y + \ell y R)}{a} \, 2 \, \df t \, \df \gf \bigg) + \sr{H_1 H_2} \lt( \fr{r^2 + y^2}{R} \, \df r^2 + \fr{r^2 + y^2}{Y} \, \df y^2 \rt) , \nnr
A_{(1)}^1 & = & \fr{2 m r s_1}{H_1 (r^2 + y^2)} \bigg( c_1 \wtd{c}_2 (1 - g^2 y^2) \, \fr{\df t}{\Xi} - \fr{c_2 \wtd{c}_1 (a^2 - y^2)}{a} \, \fr{\df \gf}{\Xi} \bigg) \nnr
&& + \fr{2 \ell y s_1}{H_1 (r^2 + y^2)} \bigg( c_1 \wtd{c}_2 (1 + g^2 r^2) \, \fr{\df t}{\Xi} - \fr{c_2 \wtd{c}_1 (r^2 + a^2)}{a} \, \fr{\df \gf}{\Xi} \bigg) , \nnr
A_{(1)}^2 & = & \fr{2 m r s_2}{H_2 (r^2 + y^2)} \bigg( c_2 \wtd{c}_1 (1 - g^2 y^2) \, \fr{\df t}{\Xi} - \fr{c_1 \wtd{c}_2 (a^2 - y^2)}{a} \, \fr{\df \gf}{\Xi} \bigg) \nnr
&& + \fr{2 \ell y s_2}{H_2 (r^2 + y^2)} \bigg( c_2 \wtd{c}_1 (1 + g^2 r^2) \, \fr{\df t}{\Xi} - \fr{c_1 \wtd{c}_2 (r^2 + a^2)}{a} \, \fr{\df \gf}{\Xi} \bigg) , \nnr
X_1 & = & H_1^{-3/4} H_2^{1/4} , \qd X_2 = H_1^{1/4} H_2^{-3/4} , \qd \chi = \fr{2 (m y - \ell r) s_1 s_2}{r^2 + y^2} , \lbl{NUTsol}
\eea
%%%
where
%%%
\bea
&& R = r^2 + a^2 - 2 m r + g^2 r^2 r_1 r_2 + a^2 g^2 r^2 - 2 m r a^2 g^2 s_1^2 s_2^2 , \qd r_I = r + 2 m s_I^2 , \nnr
&& Y = a^2 - y^2 + 2 \ell y + g^2 y^2 y_1 y_2 - a^2 g^2 y^2 + 2 \ell y a^2 g^2 s_1^2 s_2^2 , \qd y_I = y + 2 \ell s_I^2 , \nnr
&& V_r^2 = (1 + g^2 r r_1) (1 + g^2 r r_2) , \qd \wtd{V}_r^2 = (1 + r r_1 / a^2) (1 + r r_2 / a^2) , \nnr
&& V_y^2 = (1 - g^2 y y_1) (1 - g^2 y y_2) , \qd \wtd{V}_y^2 = (1 - y y_1 / a^2) (1 - y y_2 / a^2) , \nnr
&& H_I = 1 + \fr{2 (m r + \ell y) s_I^2}{r^2 + y^2} , \qd s_I = \sinh \gd_I , \qd c_I = \cosh \gd_I , \qd \wtd{c}_I = \sr{1 + a^2 g^2 s_I^2} , \nnr
&& \Xi = 1 - a^2 g^2 .
\eea
%%%
The dual potentials are\footnote{These differ by exact forms from those of \cite{chcvlupo}.}
%%%
\bea
\wtd{A}_{(1)}^1 & = & \fr{2 m y s_1}{H_2 (r^2 + y^2)} \lt( c_2 \wtd{c}_1 (1 + g^2 r r_2) \fr{\df t}{\Xi} - \fr{c_1 \wtd{c}_2 (a^2 + r r_2)}{a} \fr{\df \gf}{\Xi} \rt) \nnr
&& - \fr{2 \ell r s_1}{H_2 (r^2 + y^2)} \lt( c_2 \wtd{c}_1 (1 - g^2 y y_2) \fr{\df t}{\Xi} - \fr{c_1 \wtd{c}_2 (a^2 - y y_2)}{a} \fr{\df \gf}{\Xi} \rt) , \nnr
\wtd{A}_{(1)}^2 & = & \fr{2 m y s_2}{H_1 (r^2 + y^2)} \lt( c_1 \wtd{c}_2 (1 + g^2 r r_1) \fr{\df t}{\Xi} - \fr{c_2 \wtd{c}_1 (a^2 + r r_1)}{a} \fr{\df \gf}{\Xi} \rt) \nnr
&& - \fr{2 \ell r s_2}{H_1 (r^2 + y^2)} \lt( c_1 \wtd{c}_2 (1 - g^2 y y_1) \fr{\df t}{\Xi} - \fr{c_2 \wtd{c}_1 (a^2 - y y_1)}{a} \fr{\df \gf}{\Xi} \rt) .
\eea
%%%

The coordinate change $y = a \cos \gq$ recovers Boyer--Lindquist-type coordinates.  If $a = 0$, then we cannot use the coordinate $y$, but can remain with the $\gq$ coordinate instead.  If we let $x = \im r$ and replace $m \ra - \im m$, then the solution is invariant under the simultaneous interchange of $\ell$ and $m$ and of $x$ and $y$.  The inversion symmetry \eq{inversion} persists, provided that the NUT parameter transforms as $\ell \ra \ell / a^3 g^3$.

In the equal charge case $\gd_1 = \gd_2$, the metric in particular simplifies, becoming
%%%
\bea
\ds & = & - \fr{R}{H_1 (r^2 + y^2)} \lt( (1 - g^2 y_1 y) \, \fr{\df t}{\Xi} - \fr{a^2 - y_1 y}{a} \, \fr{\df \gf}{\Xi} \rt) ^2 + \fr{H_1 (r^2 + y^2)}{R} \, \df r^2 \nnr
&& + \fr{Y}{H_1 (r^2 + y^2)} \lt( (1 + g^2 r_1 r) \, \fr{\df t}{\Xi} - \fr{r_1 r + a^2}{a} \, \fr{\df \gf}{\Xi} \rt) ^2 + \fr{H_1 (r^2 + y^2)}{Y} \, \df y^2 ,
\eea
%%%
with the various functions defined as before.  If we let
%%%
\ben
\fr{(1 + s_1^2) (1 + a^2 g^2 s_1^2)}{s_1^2} = \fr{1 + s_1'{^2}}{s_1'{^2}} , \qd m s_1^2 = m' s_1'{^2} , \qd \ell s_1^2 = \ell' s_1'{^2} , \lbl{parameters}
\een
%%%
and then drop the primes, then all the parameters of the solution, in particular $m$, $\ell$ and $\gd$, agree with the parameters originally given in \cite{chcvlupo} when the 4 charges are pairwise equal, setting one pair of charges zero.  However, note that in \cite{chcvlupo} the coordinates $t$ and $\gf$ are not canonically normalized and asymptotically static; they are related to the $t$ and $\gf$ coordinates here by a linear transformation.  For the solution with 4 charges satisfying $\gd_1 = \gd_2$ and $\gd_3 = \gd_4 = 0$ and no NUT parameter $\ell = 0$, the thermodynamical quantities were computed in \cite{cvgilupo}, and naturally agree with those computed here in \eq{thermo}, once the parameters are related via \eq{parameters}.  Correspondingly, the BPS condition \eq{BPS} also agrees\footnote{Note a typographical error in equation (5.11) of \cite{cvgilupo}, which should be $\expe{2 (\gd_1 + \gd_2)} = 1 + 2 / a g$.}.

To discover the more general solution that includes a NUT parameter, we are again helped by the structure of the various known limits, especially from the solution without any NUT parameter.  In particular, if we let $x = \im r$ and replace $m \ra - \im m$, then the solution is symmetric under the simultaneous interchange of $\ell$ and $m$ and of $x$ and $y$.

%%%%%%%%%%%%%%%%%%%%%%%%%

\sse{Dyonic gauge fields}

%%%%%%%%%%%%%%%%%%%%%%%%%

Without axions, the field equations resulting from the $\uU (1)^4$ gauged supergravity bosonic Lagrangian \eq{Lagrangian1} are invariant under the electric/magnetic duality symmetry
%%%
\ben
F_{(2)}^I \ra X_I^{-2} \star F_{(2)}^I , \qd \gvf_i \ra - \gvf_i .
\een
%%%
Including the axions, but considering only the ungauged theory, there are 3 $\SL (2, \bbR)$ global symmetries \cite{cveticetal}.  These become manifest by dualizing 2 of the 4 gauge fields in the Lagrangian \eq{Lagrangian1}.  The gauged theory introduces a potential that breaks the $\SL (2, \bbR)$ symmetries to $\uU (1)$.

Consider our truncation of the bosonic theory to 2 gauge fields in \eq{Lagrangian2}.  Without gauging, one may dualize 1 of the 2 gauge fields to recognize the theory as a reduction of pure Einstein gravity on $T^2$.  This reduction is well-known as having an $\SL (2, \bbR)$ global symmetry.  It is also a reduction of a 5-dimensional ``bosonic string theory''; see \cite{chow}, for example.  For definiteness, we choose to dualize $F_{(2)}^1$.  To obtain the dualized Lagrangian, we enforce the Bianchi identity $\df F_{(2)}^1 = 0$ by adding to the Lagrangian the term $\wtd{A}_{(1)}^1 \wedge \df F_{(2)}^1$, considering $\wtd{A}_{(1)}^1$ as a Lagrange multiplier.  Varying with respect to $A_{(1)}^1$ gives the dual field strength $\df \wtd{A}_{(1)}^1 = \wtd{F}_{(2)}^1$ of \eq{dual}.  Substituting for $F_{(2)}^1$ and then integrating by parts, we have the dual Lagrangian
%%%
\bea
\cL & = & R \star 1 - \tf{1}{2} \star \df \gvf_1 \wedge \df \gvf_1 - \star \df \gvf_2 \wedge \df \gvf_2 - \tf{1}{2} \expe{- \gvf_1 - 2 \gvf_2} \star (\wtd{F}_{(2)}^1 + \chi F_{(2)}^2) \wedge (\wtd{F}_{(2)}^1 + \chi F_{(2)}^2) \nnr
&& - \tf{1}{2} \expe{\gvf_1 - 2 \gvf_2} \star F_{(2)}^2 \wedge F_{(2)}^2 - V \star 1 .
\eea
%%%

The global symmetries of the Lagrangian become manifest when we write it as
%%%
\ben
\cL = [R + \tf{1}{4} \Tr (\pd \cM^{-1} \, \pd \cM) - (\pd \gvf_2)^2 - \tf{1}{4} \expe{- 2 \gvf_2} \cF \tp \cM \cF - V] \star 1 ,
\een
%%%
where
%%%
\ben
\cM = \expe{- \gvf_1}
\begin{pmatrix}
\chi^2 + \expe{2 \gvf_1} & \chi \\
\chi & 1
\end{pmatrix} , \qd
\cF =
\begin{pmatrix}
F_{(2)}^2 \\
\wtd{F}_{(2)}^1
\end{pmatrix} =
\begin{pmatrix}
\df A_{(1)}^2 \\
\df \wtd{A}_{(1)}^1
\end{pmatrix} ,
\een
%%%
in terms of which the potential is $V = - g^2 (\Tr \cM + 4 \cosh \gvf_2)$.  Without any potential, the Lagrangian has a manifest global $\SL (2, \bbR)$ symmetry under
%%%
\ben
\cF \ra (\gL \tp )^{-1} \cF , \qd \cM \ra \gL \cM \gL \tp , \qd \gL \in \SL (2, \bbR) .
\een
%%%
This implies that the potentials and dual potentials transform as
%%%
\ben
\begin{pmatrix}
A_{(1)}^2 \\
\wtd{A}_{(1)}^1
\end{pmatrix}
\ra
(\gL \tp)^{-1}
\begin{pmatrix}
A_{(1)}^2 \\
\wtd{A}_{(1)}^1
\end{pmatrix}
, \qd
\begin{pmatrix}
A_{(1)}^1 \\
\wtd{A}_{(1)}^2
\end{pmatrix}
\ra
(\gL \tp)^{-1}
\begin{pmatrix}
A_{(1)}^1 \\
\wtd{A}_{(1)}^2
\end{pmatrix} ; \lbl{Atransform}
\een
%%%
these two sets of transformation rules are equivalent once the scalar transformation rules are taken into account.  More explicitly, if
%%%
\ben
\gL =
\begin{pmatrix}
a & b \\
c & d
\end{pmatrix} , \qd a d - b c = 1 ,
\een
%%%
then the axidilaton complex scalar $\gt = \chi + \im \expe{\gvf_1}$ transforms as
%%%
\ben
\gt \ra \fr{a \gt + b}{c \gt + d} ,
\een
%%%
i.e.~the dilaton and axion transform as
%%%
\bea
\expe{- \gvf_1} & \ra & \expe{- \gvf_1} [(c \chi + d)^2 + c^2 \expe{2 \gvf_1}] , \nnr
\chi & \ra & \fr{(a \chi + b) (c \chi + d) + a c \expe{2 \gvf_1}}{(c \chi + d)^2 + c^2 \expe{2 \gvf_1}} .
\eea
%%%

When we include gauging, the potential breaks the $\SL (2, \bbR)$ global symmetry to $\uU (1)$, with $\gL$ restricted to the form
%%%
\ben
\gL =
\begin{pmatrix}
\cos \ga & - \sin \ga \\
\sin \ga & \cos \ga 
\end{pmatrix} .
\een
%%%
This preserves the vanishing of $\gvf_1$ and $\chi$ as $r \ra \infty$.  We can apply the symmetry to transform our general solution \eq{NUTsol}.  The metric and one scalar $\expe{\gvf_2} = \sr{H_1 / H_2}$ remain the same, the gauge potentials are simply rotated according to \eq{Atransform}, and the remaining scalars become
%%%
\bea
\chi & = & \fr{\cos (2 \ga) \, 2 (m y - \ell r) s_1 s_2 + \tf{1}{2} \sin (2 \ga) (r_1 r_2 + y_1 y_2 - r^2 - y^2)}{\cos^2 \ga \, (r^2 + y^2) + \sin^2 \ga \, (r_1 r_2 + y_1 y_2) + \sin (2 \ga) \, 2 (m y - \ell r) s_1 s_2} , \nnr
\expe{\gvf_1} & = & \fr{\sr{H_1 H_2} (r^2 + y^2)}{\cos^2 \ga \, (r^2 + y^2) + \sin^2 \ga \, (r_1 r_2 + y_1 y_2) + \sin (2 \ga) \, 2 (m y - \ell r) s_1 s_2} .
\eea
%%%

For ungauged supergravity, it was shown in \cite{sen} how to construct dyonic Kerr(--NUT) solutions with a metric involving 2 charge parameters.  For the simpler ungauged case corresponding to our 2 charge parameters being equal, see \cite{galkec}.

%%%%%%%%%%%%%%%%%%%%%%%%%%%%%%%%%%%%

\se{Killing tensor and separability}

%%%%%%%%%%%%%%%%%%%%%%%%%%%%%%%%%%%%

A rank-2 Killing--St\"{a}ckel tensor is a symmetric tensor $K_{a b}$ that satisfies $\na_{( a} K_{b c )} = 0$.  Charged and rotating supergravity black hole solutions that generalize the Kerr or Myers--Perry solution generally seem to have string frame metrics admit such tensors, and consequently geodesic motion in the string frame metric is completely integrable \cite{chow}.  Furthermore the massless Klein--Gordon equation separates for the Einstein frame metric.  We shall see that these properties continue to hold for our 2-charge solution.

%%%%%%%%%%%%%%%%%%%%

\sse{Killing tensor}

%%%%%%%%%%%%%%%%%%%%

The string frame metric $\df \wtd{s}^2$ is related to the original Einstein frame metric $\ds$ by $\ds = \sr{H_1 H_2} \, \df \wtd{s}^2$.  The inverse string frame metric is, including the NUT parameter,
%%%
\bea
\lt( \fr{\pd}{\pd \wtd{s}} \rt) ^2 = \wtd{g}^{a b} \, \pd_a \, \pd_b & = & - \fr{a^4}{r^2 + y^2} \bigg( \fr{\wtd{V}_r^2}{R} - \fr{\wtd{V}_y^2}{Y} \bigg) \, \pd_t^2 - \fr{2 c_1 c_2 \wtd{c}_1 \wtd{c}_2 a}{r^2 + y^2} \lt( \fr{m r}{R} + \fr{\ell y}{Y} \rt) \, 2 \, \pd_t \, \pd_\gf \nnr
&& + \fr{1}{r^2 + y^2} \lt( \fr{V_y^2}{Y} - \fr{V_r^2}{R} \rt) a^2 \, \pd_\gf^2 + \fr{R}{r^2 + y^2} \, \pd_r^2 + \fr{Y}{r^2 + y^2} \, \pd_y^2 .
\eea
%%%
The components $(r^2 + y^2) \wtd{g}^{a b}$ are additively separable as a function of $r$ plus a function of $y$.

For the string frame metric, a rank-2 Killing--St\"{a}ckel tensor is
%%%
\ben
\wtd{K} = \wtd{K}^{a b} \, \pd_a \, \pd_b = \fr{a^4 \wtd{V}_y^2}{Y} \, \pd_t^2 - \fr{2 c_1 c_2 \wtd{c}_1 \wtd{c}_2 \ell y a}{Y} \, 2 \, \pd_t \, \pd_\gf + \fr{V_y^2 a^2}{Y} \, \pd_\gf^2 + Y \, \pd_y^2 - y^2 \lt( \fr{\pd}{\pd \wtd{s}} \rt) ^2 .
\een
%%%
It is irreducible, i.e.~is not merely a linear combination of the metric and outer products of Killing vectors.

A rank-2 conformal Killing--St\"{a}ckel tensor is a symmetric tensor $Q_{a b}$ that satisfies an equation of the form $\na_{( a} Q_{b c)} = q_{(a } g_{b c)}$, for some $q_a$.  In 4 dimensions, we can express $q_a$ in terms of $Q_{a b}$ as $q_a = \tf{1}{6} (\pd_a Q{^b}{_b} + 2 \na_b Q{^b}{_a})$.  There is an induced rank-2 conformal Killing--St\"{a}ckel tensor, with components $Q^{a b} = \wtd{K}^{a b}$, for any conformally related metric, in particular the Einstein frame metric.  If $q_a$ is a gradient, i.e.~$q_a = \pd_a q$ for some scalar $q$, then the rank-2 conformal Killing--St\"{a}ckel tensor $Q_{a b}$ is said to be of gradient type.  In this case, one may then construct an associated Killing--St\"{a}ckel tensor $K_{a b} = Q_{a b} - q g_{a b}$.  Here, the problem of determining whether the induced $Q_{a b}$ for the Einstein frame metric is of gradient type reduces to computing
%%%
\ben
\pd_r q_y - \pd_y q_r = \fr{(s_1^2 - s_2^2)^2 (m r^2 + 2 \ell r y - m y^2) (\ell r^2 - 2 m r y - \ell y^2)}{H_1^{3/2} H_2^{3/2} (r^2 + y^2)^3} .
\een
%%%
Therefore, for two equal charges, $\gd_1 = \gd_2$, the Einstein frame metric possesses an irreducible rank-2 Killing--St\"{a}ckel tensor.  It can be obtained by finding that $q_a = \pd_a q$, with
%%%
\ben
q = \fr{2 (m r + \ell y) s_1^2 y^2}{H_1 (r^2 + y^2)} + 2 \ell s_1^2 y .
\een
%%%
Such a tensor is known \cite{vasudevan, chow}, and agrees, up to a reducible Killing--St\"{a}ckel tensor, with that constructed here.

%%%%%%%%%%%%%%%%%%

\sse{Separability}

%%%%%%%%%%%%%%%%%%

Because the string frame metric possesses a rank-2 Killing--St\"{a}ckel tensor and furthermore satisfies the properties required for a separability structure, geodesic motion in the string frame metric is completely integrable.  We shall demonstrate this explicitly, and also demonstrate the separability of the Einstein frame massless Klein--Gordon equation.  Note that without rotation, for which our $y$ coordinate cannot be used, the enhanced isometry group guarantees separation of these equations.

%%%%%%%%%%%%%%%%%%%%%%%%%%%%%%%%

\ssse{Hamilton--Jacobi equation}

%%%%%%%%%%%%%%%%%%%%%%%%%%%%%%%%

The Hamiltonian for geodesic motion in the string frame metric is
%%%
\ben
H (x^a , p_b) = \tf{1}{2} \wtd{g}^{a b} p_a p_b ,
\een
%%%
where $p_a$ are the momenta conjugate to $x^a$.  Therefore, the Hamilton--Jacobi equation is
%%%
\ben
\fr{\pd S}{\pd \gl} + \fr{1}{2} \wtd{g}^{a b} \, \pd_a S \, \pd_b S = 0 ,
\een
%%%
where $S$ is Hamilton's principal function, with $\pd_a S = p_a = \df x_a / \df \gl$, and $\gl$ is an affine parameter along the worldline of a particle.  Consider the separable ansatz
%%%
\ben
S = \fr{1}{2} \mu^2 \gl - E t + L \gf + S_r (r) + S_y (y) ,
\een
%%%
where the constants $E$ and $L$ are momenta conjugate to the ignorable coordinates $t$ and $\gf$, representing conserved energy and angular momentum respectively.  $\mu$ is the mass of the particle, satisfying $p^a p_a = - \mu^2$.  We therefore have
%%%
\bea
&& - a^4 \bigg( \fr{\wtd{V}_r^2}{R} - \fr{\wtd{V}_y^2}{Y} \bigg) E^2 + 4 c_1 c_2 \wtd{c}_1 \wtd{c}_2 a \lt( \fr{m r}{R} + \fr{\ell y}{Y} \rt) E L + \lt( \fr{V_y^2}{Y} - \fr{V_r^2}{R} \rt) a^2 L^2 \nnr
&& + R \lt( \fr{\df S_r}{\df r} \rt) ^2 + Y \lt( \fr{\df S_y}{\df y} \rt) ^2 + \mu^2 (r^2 + y^2) = 0 , \lbl{HJ}
\eea
%%%
which is additively separable.  Introducing a separation constant $C$, we have
%%%
\bea
\fr{\df r}{\df \gl} & = & \wtd{g}^{r r} p_r = \fr{R}{r^2 + y^2} \fr{\df S_r}{\df r} , \nnr
\fr{\df y}{\df \gl} & = & \wtd{g}^{y y} p_y = \fr{Y}{r^2 + y^2} \fr{\df S_y}{\df y} ,
\eea
%%%
where
%%%
\bea
S_r & = & \int \! \df r \, \fr{1}{R} \sr{\wtd{V}_r^2 a^4 E^2 - 4 c_1 c_2 \wtd{c}_1 \wtd{c}_2 m r a E L + V_r^2 a^2 L^2 - \mu^2 r^2 R - C R} , \nnr
S_y & = & \int \! \df y \, \fr{1}{Y} \sr{- \wtd{V}_y^2 a^4 E^2 - 4 c_1 c_2 \wtd{c}_1 \wtd{c}_2 \ell y a E L - V_y^2 a^2 L^2 - \mu^2 y^2 Y + C Y} ,
\eea
%%%
which determines $r(\gl)$ and $y(\gl)$ by quadratures.  We then have
%%%
\bea
\fr{\df t}{\df \gl} & = & \wtd{g}^{t t} p_t + \wtd{g}^{t \gf} p_\gf = \fr{a^4}{r^2 + y^2} \bigg( \fr{\wtd{V}_r^2}{R} - \fr{\wtd{V}_y^2}{Y} \bigg) E - \fr{2 c_1 c_2 \wtd{c}_1 \wtd{c}_2 a}{r^2 + y^2} \lt( \fr{m r}{R} + \fr{\ell y}{Y} \rt) L , \nnr
\fr{\df \gf}{\df \gl} & = & \wtd{g}^{t \gf} p_t + \wtd{g}^{\gf \gf} p_\gf = \fr{2 c_1 c_2 \wtd{c}_1 \wtd{c}_2 a}{r^2 + y^2} \lt( \fr{m r}{R} + \fr{\ell y}{Y} \rt) E + \fr{1}{r^2 + y^2} \lt( \fr{V_y^2}{Y} - \fr{V_r^2}{R} \rt) a^2 L ,
\eea
%%%
which determines $t(\gl)$ and $\gf(\gl)$ by quadratures.

For the Einstein frame metric, we can follow the same procedure.  The only difference is that, in \eq{HJ}, we replace $\mu^2$ by $\sr{H_1 H_2} \mu^2$.  Therefore, separation only occurs for null geodesics or if the two charges are equal.

%%%%%%%%%%%%%%%%%%%%%%%%%%%%%

\ssse{Klein--Gordon equation}

%%%%%%%%%%%%%%%%%%%%%%%%%%%%%

The (minimally coupled) Klein--Gordon equation for the Einstein frame metric is
%%%
\ben
\square \gF = \fr{1}{\sr{-g}} \pd_a (\sr{-g} g^{a b} \pd_b \gF) = \mu^2 \gF .
\een
%%%
Consider the separable ansatz
%%%
\ben
\gF = \gF_r (r) \gF_y (y) \expe{\im (k \gf - \gw t)} ,
\een
%%%
and note that
%%%
\ben
\sr{-g} = \fr{\sr{H_1 H_2} (r^2 + y^2)}{a \Xi} .
\een
%%%
We therefore have
%%%
\bea
\sr{H_1 H_2} (r^2 + y^2) \mu^2 & = & - a^4 \bigg( \fr{\wtd{V}_r^2}{R} - \fr{\wtd{V}_y^2}{Y} \bigg) \gw^2 - 4 c_1 c_2 \wtd{c}_1 \wtd{c}_2 a \lt( \fr{m r}{R} + \fr{\ell y}{Y} \rt) \gw k - \lt( \fr{V_y^2}{Y} - \fr{V_r^2}{R} \rt) k^2 \nnr
&& + \fr{1}{\gF_r} \fr{\df}{\df r} \lt( R \fr{\df \gF_r}{\df r} \rt) + \fr{1}{\gF_y} \fr{\df}{\df y} \lt( Y \fr{\df \gF_y}{\df y} \rt) .
\eea
%%%
In general, only the massless Klein--Gordon equation, with $\mu = 0$, is separable, but in the special case of two equal charges, with $\gd_1 = \gd_2$, the massive equation is separable \cite{vasudevan}.  The separated massless equations are
%%%
\bea
&& \fr{1}{R} \fr{\df}{\df r} \lt( R \fr{\df \gF_r}{\df r} \rt) + \fr{- a^4 \wtd{V}_r^2 \gw^2 - 4 c_1 c_2 \wtd{c}_1 \wtd{c}_2 a m r \gw k + V_r^2 k^2}{R^2} \gF_r = 0 , \nnr
&& \fr{1}{Y} \fr{\df}{\df y} \lt( Y \fr{\df \gF_y}{\df y} \rt) + \fr{a^4 \wtd{V}_y^2 \gw^2 - 4 c_1 c_2 \wtd{c}_1 \wtd{c}_2 a \ell y \gw k - V_y^2 k^2}{Y^2} \gF_y = 0 .
\eea
%%%
In general, these are Fuchsian second order ordinary differential equations with 5 regular singular points in the complex $r$ or $y$ planes, including one at infinity.  However, following \cite{sutaum}, we expect that it is possible to perform a transformation to factor out 1 singularity.  4 regular singular points would remain, reducing the differential equations to the Heun equation.

%%%%%%%%%%%%%%%%%%%%

\section{Conclusion}

%%%%%%%%%%%%%%%%%%%%

We have presented an asymptotically AdS rotating black hole solution of 4-dimensional $\uU (1)^4$ gauged supergravity that possesses 2 non-zero $\uU (1)$ charges, and studied some of its properties.  Although there are BPS solutions, they cannot be black holes, but instead are nakedly singular.  It would be interesting to include more charges, in order to find further AdS$_4$ black holes that are supersymmetric.  We also obtained more general solutions with a NUT parameter and allowed the gauge fields to be dyonic.  Another parameter that we have not attempted to include is an acceleration parameter, which would provide generalizations of the C-metric; such a generalization is known in the ungauged case \cite{chcvlupo}.  As expected, there are hidden symmetries, with the string frame metric admitting a rank-2 Killing--St\"{a}ckel tensor.  It would be interesting to further investigate hidden symmetries of the solution along the lines of \cite{hokuwaya}.

%%%%%%%%%%%%%%%%%%%%%%%%%%

%\section*{Acknowledgement}

%%%%%%%%%%%%%%%%%%%%%%%%%%


\begin{thebibliography}{99}

%\cite{Emparan:2008eg}
\bibitem{emprea}
  R.~Emparan and H.S.~Reall,
  ``Black holes in higher dimensions,''
  \textit{Living Rev.\ Rel.} {\bf 11}, 6 (2008)
  [\texttt{arXiv:0801.3471}].
  %%CITATION = 00222,11,6;%%

%\cite{Chow:2008fe}
\bibitem{chow}
  D.D.K.~Chow,
  ``Symmetries of supergravity black holes,''
  \textit{Class.\ Quant.\ Grav.} {\bf 27}, 205009 (2010)
  [\texttt{arXiv:0811.1264}].
  %%CITATION = CQGRD,27,205009;%%

%\cite{Chow:2010sf}
\bibitem{chow0}
  D.D.K.~Chow,
  ``Single-charge rotating black holes in four-dimensional gauged
  supergravity,''
  \textit{Class.\ Quant.\ Grav.} {\bf 28}, 032001 (2011)
  \texttt{arXiv:1011.2202}.
  %%CITATION = ARXIV:1011.2202;%%

\bibitem{carter0}
B.~Carter,
``A new family of Einstein spaces,''
\textit{Phys.\ Lett.\ A} {\bf 26}, 399 (1968).

%\cite{Carter:1968ks}
\bibitem{carter}
  B.~Carter,
  ``Hamilton--Jacobi and Schr\"{o}dinger separable solutions of Einstein's
  equations,''
  \textit{Commun.\ Math.\ Phys.} {\bf 10}, 280 (1968).
  %%CITATION = CMPHA,10,280;%

%\cite{Chong:2004na}
\bibitem{chcvlupo}
  Z.-W.~Chong, M.~Cveti\v{c}, H.~L\"{u} and C.N.~Pope,
  ``Charged rotating black holes in four-dimensional gauged and ungauged
  supergravities,''
  \textit{Nucl.\ Phys.\  B} {\bf 717}, 246 (2005)
  [\texttt{hep-th/0411045}].
  %%CITATION = NUPHA,B717,246;%%

%\cite{Cvetic:1996dt}
\bibitem{cveyou3}
  M.~Cveti\v{c} and D.~Youm,
  ``Near-BPS-saturated rotating electrically charged black holes as string
  states,''
  \textit{Nucl.\ Phys.\  B} {\bf 477}, 449 (1996)
  [\texttt{hep-th/9605051}].
  %%CITATION = NUPHA,B477,449;%%

%\cite{Chow:2007ts}
\bibitem{chow3}
  D.D.K.~Chow,
  ``Equal charge black holes and seven-dimensional gauged supergravity,''
  \textit{Class.\ Quant.\ Grav.}  {\bf 25}, 175010 (2008)
  [\texttt{arXiv:0711.1975}].
  %%CITATION = CQGRD,25,175010;%%

%\cite{Chong:2005da}
\bibitem{chcvlupo2}
  Z.-W.~Chong, M.~Cveti\v{c}, H.~L\"{u} and C.N.~Pope,
  ``Five-dimensional gauged supergravity black holes with independent rotation
  parameters,''
  \textit{Phys.\ Rev.\  D} {\bf 72}, 041901 (2005)
  [\texttt{hep-th/0505112}].
  %%CITATION = PHRVA,D72,041901;%%

%\cite{Chow:2008ip}
\bibitem{chow2}
  D.D.K.~Chow,
  ``Charged rotating black holes in six-dimensional gauged supergravity,''
  \textit{Class.\ Quant.\ Grav.} {\bf 27}, 065004 (2010)
  [\texttt{arXiv:0808.2728}].
  %%CITATION = CQGRD,27,065004;%%

%\cite{Cvetic:1999xp}
\bibitem{cveticetal}
  M.~Cveti\v{c} {\it et al.},
  ``Embedding AdS black holes in ten and eleven dimensions,''
  \textit{Nucl.\ Phys.\  B} {\bf 558}, 96 (1999)
  [\texttt{hep-th/9903214}].
  %%CITATION = NUPHA,B558,96;%%

%\cite{Duff:1999gh}
\bibitem{dufliu}
  M.J.~Duff and J.T.~Liu,
  ``Anti-de Sitter black holes in gauged $N = 8$ supergravity,''
  \textit{Nucl.\ Phys.\  B} {\bf 554}, 237 (1999)
  [\texttt{hep-th/9901149}].
  %%CITATION = NUPHA,B554,237;%%

%\cite{Sen:1994eb}
\bibitem{sen}
  A.~Sen,
  ``Black hole solutions in heterotic string theory on a torus,''
  \textit{Nucl.\ Phys.\  B} {\bf 440}, 421 (1995)
  [\texttt{hep-th/9411187}].
  %%CITATION = NUPHA,B440,421;%%

%\cite{Cvetic:1996kv}
\bibitem{cveyou}
  M.~Cveti\v{c} and D.~Youm,
  ``Entropy of nonextreme charged rotating black holes in string theory,''
  \textit{Phys.\ Rev.\ D} {\bf 54}, 2612 (1996)
  [\texttt{hep-th/9603147}].
  %%CITATION = PHRVA,D54,2612;%%

%\cite{Jatkar:1996kd}
\bibitem{jamupa}
  D.P.~Jatkar, S.~Mukherji and S.~Panda,
  ``Rotating dyonic black holes in heterotic string theory,''
  \textit{Phys.\ Lett.\  B} {\bf 384}, 63 (1996)
  [\texttt{hep-th/9601118}].
  %%CITATION = PHLTA,B384,63;%%

%\cite{Frolov:1987rj}
\bibitem{frzebl}
  V.P.~Frolov, A.I.~Zelnikov and U.~Bleyer,
  ``Charged rotating black hole from five-dimensional point of view,''
  \textit{Annalen Phys.}  {\bf 44}, 371 (1987).
  
%\cite{Chen:2006ea}
\bibitem{chlupo3}
  W.~Chen, H.~L\"{u} and C.N.~Pope,
  ``Kerr--de Sitter black holes with NUT charges,''
  \textit{Nucl.\ Phys.\  B} {\bf 762}, 38 (2007)
  [\texttt{hep-th/0601002}].
  %%CITATION = NUPHA,B762,38;%%

%\cite{Hawking:1998kw}
\bibitem{hahuta}
  S.W.~Hawking, C.J.~Hunter and M.M.~Taylor-Robinson,
  ``Rotation and the AdS--CFT correspondence,''
  \textit{Phys.\ Rev.\  D} {\bf 59}, 064005 (1999)
  [\texttt{hep-th/9811056}].
  %%CITATION = PHRVA,D59,064005;%%

%\cite{Chen:2006xh}
\bibitem{chlupo2}
  W.~Chen, H.~L\"{u} and C.N.~Pope,
  ``General Kerr--NUT--AdS metrics in all dimensions,''
  \textit{Class.\ Quant.\ Grav.}  {\bf 23}, 5323 (2006)
  [\texttt{hep-th/0604125}].
  %%CITATION = CQGRD,23,5323;%%

%\cite{de Wit:1986iy}
\bibitem{dewnic}
  B.~de Wit and H.~Nicolai,
  ``The consistency of the $S^7$ truncation in $d = 11$ supergravity,''
  \textit{Nucl.\ Phys.\  B} {\bf 281}, 211 (1987).
  %%CITATION = NUPHA,B281,211;%%

%\cite{Cvetic:1999au}
\bibitem{cvlupo}
  M.~Cveti\v{c}, H.~L\"{u} and C.N.~Pope,
  ``Four-dimensional $N = 4$, $\SO(4)$ gauged supergravity from $D = 11$,''
  \textit{Nucl.\ Phys.\  B} {\bf 574}, 761 (2000)
  [\texttt{hep-th/9910252}].
  %%CITATION = NUPHA,B574,761;%%

%\cite{Gibbons:2004ai}
\bibitem{gipepo}
  G.W.~Gibbons, M.J.~Perry and C.N.~Pope,
  ``The first law of thermodynamics for Kerr--anti-de Sitter black holes,''
  \textit{Class.\ Quant.\ Grav.}  {\bf 22}, 1503 (2005)
  [\texttt{hep-th/0408217}].
  %%CITATION = CQGRD,22,1503;%%

%\cite{Ashtekar:1984zz}
\bibitem{ashmag}
  A.~Ashtekar and A.~Magnon,
  ``Asymptotically anti-de Sitter space--times,''
  \textit{Class.\ Quant.\ Grav.}  {\bf 1} (1984) L39.
  %%CITATION = CQGRD,1,L39;%%

%\cite{Ashtekar:1999jx}
\bibitem{ashdas}
  A.~Ashtekar and S.~Das,
  ``Asymptotically anti-de Sitter space-times: conserved quantities,''
  \textit{Class.\ Quant.\ Grav.}  {\bf 17}, L17 (2000)
  [\texttt{hep-th/9911230}].
  %%CITATION = CQGRD,17,L17;%%

%\cite{Chen:2005zj}
\bibitem{chlupo}
  W.~Chen, H.~L\"{u} and C.N.~Pope,
  ``Mass of rotating black holes in gauged supergravities,''
  \textit{Phys.\ Rev.\  D} {\bf 73}, 104036 (2006)
  [\texttt{hep-th/0510081}].
  %%CITATION = PHRVA,D73,104036;%%

%\cite{Kostelecky:1995ei}
\bibitem{kosper}
  V.A.~Kosteleck\'{y} and M.J.~Perry,
  ``Solitonic black holes in gauged $N = 2$ supergravity,''
  \textit{Phys.\ Lett.\  B} {\bf 371}, 191 (1996)
  [\texttt{hep-th/9512222}].
  %%CITATION = PHLTA,B371,191;%%
  
%\cite{Caldarelli:1998hg}
\bibitem{calkle}
  M.M.~Caldarelli and D.~Klemm,
  ``Supersymmetry of anti-de Sitter black holes,''
  \textit{Nucl.\ Phys.\  B} {\bf 545}, 434 (1999)
  [\texttt{hep-th/9808097}].
  %%CITATION = NUPHA,B545,434;%%

%\cite{Cvetic:2005zi}
\bibitem{cvgilupo}
  M.~Cveti\v{c}, G.W.~Gibbons, H.~L\"{u} and C.N.~Pope,
  ``Rotating black holes in gauged supergravities; thermodynamics,
  supersymmetric limits, topological solitons and time machines,''
  \texttt{hep-th/0504080}.
  %%CITATION = HEP-TH/0504080;%%

%\cite{Kunduri:2006ek}
\bibitem{kulure}
  H.K.~Kunduri, J.~Lucietti and H.S.~Reall,
  ``Supersymmetric multi-charge AdS$_5$ black holes,''
  \textit{JHEP} {\bf 0604}, 036 (2006)
  [\texttt{hep-th/0601156}].
  %%CITATION = JHEPA,0604,036;%%

%\cite{Galtsov:1994pd}
\bibitem{galkec}
  D.V.~Gal'tsov and O.V.~Kechkin,
  ``Ehlers--Harrison-type transformations in dilaton-axion gravity,''
  \textit{Phys.\ Rev.\  D} {\bf 50}, 7394 (1994)
  [\texttt{hep-th/9407155}].
  %%CITATION = PHRVA,D50,7394;%%

%\cite{Vasudevan:2005bz}
\bibitem{vasudevan}
  M.~Vasudevan,
  ``Integrability of some charged rotating supergravity black hole solutions
  in four and five dimensions,''
  \textit{Phys.\ Lett.\  B} {\bf 624}, 287 (2005)
  [\texttt{gr-qc/0507092}].
  %%CITATION = PHLTA,B624,287;%%

%\cite{Suzuki:1998vy}
\bibitem{sutaum}
  H.~Suzuki, E.~Takasugi and H.~Umetsu,
  ``Perturbations of Kerr--de Sitter black holes and Heun's equations,''
  \textit{Prog.\ Theor.\ Phys.} {\bf 100}, 491 (1998)
  [\texttt{gr-qc/9805064}].
  %%CITATION = PTPKA,100,491;%%

%\cite{Houri:2010fr}
\bibitem{hokuwaya}
  T.~Houri, D.~Kubiz\v{n}\'{a}k, C.M.~Warnick and Y.~Yasui,
  ``Generalized hidden symmetries and the Kerr--Sen black hole,''
  \textit{JHEP} {\bf 1007}, 055 (2010)
  [\texttt{arXiv:1004.1032}].
  %%CITATION = JHEPA,1007,055;%%

\end{thebibliography}
\end{document}